\documentclass[11pt,a4paper]{article}
\pdfoutput=1
\usepackage{jheppub1}
\usepackage{amsfonts,amsmath,amssymb}
\usepackage{enumerate}
\usepackage{hyperref}
\usepackage{graphicx}
\usepackage{caption}
\usepackage{subcaption}

\usepackage{enumitem}

\newcommand{\be}{\begin{equation}}
\newcommand{\ee}{\end{equation}}

\newcommand{\bea}{\begin{eqnarray}}
\newcommand{\eea}{\end{eqnarray}}

\newcommand{\bes}{\begin{subequations}}
\newcommand{\ees}{\end{subequations}}

\newcommand{\nn}{\nonumber}

\title{Black holes with baryonic charge and $\mathcal{I}$-extremization}
\author[a]{Hyojoong Kim}
\author[a,b]{and Nakwoo Kim}

\affiliation[a]{Department of Physics 
and Research Institute of Basic Science, \\ Kyung Hee University, 
Seoul 02447, Republic of Korea}
\affiliation[a]{School of Physics, Korea Institute for Advanced Study, Seoul 02445, Republic of Korea}

\emailAdd{h.kim@khu.ac.kr}
\emailAdd{nkim@khu.ac.kr}

\abstract{Recently it was discovered that twisted superconformal index ${\mathcal{I}}$ can be used to understand the Bekenstein-Hawking entropy of magnetically charged black holes in AdS spacetime. In this paper we apply the so-called $\mathcal{I}$-extremization procedure to three-dimensional gauge field theories and their geometric dual, focusing in particular on the seven-dimensional Sasaki-Einstein manifold $M^{1,1,1}$. We generalize recent studies on relations among toric geometry, variational principles, and black hole entropy to the case of AdS$_2 \times Y_9$, where $Y_9$ is a fibration of toric Sasaki-Einstein manifold $M^{1,1,1}$ over a two-dimensional Riemann surface $\Sigma_g$. The nine-dimensional variational problem is given in terms of an entropy functional. In order to illustrate the computations as explicitly as possible, we consider cases where either only mesonic or baryonic fluxes are turned on. 
By employing the operator counting method, we calculate the $S^3$ free energy and the topologically twisted index $\mathcal{I}$ at large-$N$. The result for $\mathcal{I}$, it turns out, can be also obtained from the variational principle of the entropy functional with mesonic fluxes. We also study asymptotically AdS${}_4$ black holes which are magnetically charged with 
respect to the vector field in the Betti multiplet. By extremizing the entropy functional with baryonic flux, we compute the entropy and find that it agrees with the entropy of an explicit solution in a four-dimensional gauged supergravity which is a consistent truncation of eleven-dimensional supergravity in AdS${}_4\times M^{1,1,1}$.}

\begin{document}
\maketitle
\flushbottom
\section{Introduction}
Microscopic understanding of the black hole entropy has been one of the most important themes of string theory over the past decades, ever since Strominger and Vafa's successful work on asymptotically flat black holes \cite{Strominger:1996sh}. From the viewpoint of AdS/CFT correspondence \cite{Maldacena:1997re}, it is also interesting to study asymptotically AdS black holes.
Static and supersymmetric AdS black holes with spherical horizon were constructed {\it e.g.} in \cite{Cacciatori:2009iz, DallAgata:2010ejj,Hristov:2010ri}. In these works the authors considered magnetically charged solutions in $D=4,\, \mathcal{N}=2$ gauged supergravity coupled to three vector multiplets, which we refer to as the STU model. The AdS/CFT dual is the ABJM theory \cite{Aharony:2008ug}, put on $S^1 \times S^2$ with a topological twist on $S^2$. Following the development of the localization technique \cite{Pestun:2007rz}, the relevant topologically twisted index was calculated in \cite{Benini:2015noa,Benini:2016hjo}, and it was shown that the results agree with the entropy of magnetically charged black holes \cite{Benini:2015eyy, Benini:2016rke}. After more than 20 years since Strominger and Vafa's celebrated work, we have the first satisfactory microscopic understanding of asymptotically AdS black holes. This triumph has led to intensive studies on this subject. Another direction of recent years is the calculation of the entropy for rotating, electrically charged black holes using dual field theory, in \cite{Cabo-Bizet:2018ehj, Choi:2018hmj, Benini:2018ywd}. For a review and more complete set of references, see for example \cite{Zaffaroni:2019dhb}. 

The topologically twisted index is a function of the chemical potential and the magnetic flux. Then, the index is extremized with respect to the chemical potential. This procedure is called $\mathcal{I}$-extremization \cite{Benini:2015eyy, Benini:2016rke}. Its gravity dual is known for AdS$_4$ black holes in $\mathcal{N}=2$ gauged supergravity \cite{Cacciatori:2009iz, DallAgata:2010ejj,Hristov:2010ri}.
The BPS equations for the near horizon geometry can be rephrased as an extremization problem of some function of scalar fields. It is called an attractor equation in the sense that the extremization determines the values of the scalar fields on the horizon of the black hole. Then the black hole entropy  is given by its extremum.

Recently, the geometric dual of $\mathcal{I}$-extremization was proposed in \cite{Couzens:2018wnk} as a variational problem of certain {\it off-shell} configurations in supergravity.
In \cite{Couzens:2018wnk}, the authors studied the off-shell AdS$_3$  $\times$ Y$_7$ and AdS$_2$ $\times$ Y$_9$ solutions of type IIB and D=11 supergravity, respectively. Here off-shell means that one imposes the conditions for supersymmetry and relaxes the equations of motion for the five-form flux. One constructs seven- and nine-dimensional actions  $S_{\textrm{SUSY}}$ and study their variational problems. Here $S_{\textrm{SUSY}}$  depends on the R-symmetry vector field and the transverse K\"ahler class of Y$_{2n+1}$.
From AdS$_2$ $\times$ Y$_9$ solutions of D=11 supergravity,
they proposed that
the Bekenstein-Hawking entropy can be obtained by extremizing $S_{\textrm{SUSY}}$ as
\be
S_{\textrm{BH}}=\dfrac{1}{4G_2}=8\pi^2 {S_{\textrm{SUSY}}|_{\textrm{on-shell}}}.
\ee
The supersymmetric AdS$_4$ black hole they considered has an AdS$_2$ $\times$ $\Sigma_g$ near-horizon geometry and can be uplifted on a Sasaki-Einstein seven-manifold.

In a subsequent paper \cite{Gauntlett:2018dpc}, the authors focus on  AdS$_3$  $\times$ Y$_7$ solution of type IIB supergravity where a seven-dimensional manifold $Y_7$ is a fibration of toric $Y_5$ over a two-dimensional Riemann surface $\Sigma_g$ with genus $g>1$. They construct the so-called {\it master volume}, the form of which is determined by the toric data of $Y_5$. Given the master volume, which satisfy the constraint equation and the flux quantization conditions, one can easily compute the action functional and the R-charges of baryonic operators. This variational problem corresponds to a geometric dual of $c$-extremization \cite{Benini:2012cz, Benini:2013cda}. Recently it was shown that the equivalence holds to extend off-shell for all toric quivers \cite{Hosseini:2019use}.

The geometric dual of $\mathcal{I}$-extremization for ABJM theory was studied in \cite{Hosseini:2019use}. Based on the construction of \cite{Couzens:2018wnk, Gauntlett:2018dpc}, the authors studied the case where a nine-dimensional manifold $Y_9$ is a $S^7$ fibration over $\Sigma_g$. They employed the toric data of $S^7$ and obtained the master volume explicitly. Then calculating the entropy functional, they showed that it exactly reproduces the topologically twisted index of ABJM theory off-shell \cite{Benini:2015eyy, Benini:2016rke}.

In this work, we aim to generalize the discussions of \cite{Hosseini:2019use} to the seven-dimensional Sasaki-Einstein manifold $M^{1,1,1}$. Recall that $M^{1,1,1}$ is an example of homogeneous Sasaki-Einstein manifolds in seven dimensions, with isometry group $SU(3)\times SU(2)\times U(1)$.
We study the M2-branes placed at the tip of the cone over $M^{1,1,1}$ and in particular their compactification on $\Sigma_g$. Using the toric data of $M^{1,1,1}$, we construct the master volume and compute the entropy functional.

In contrast to $S^7$, $M^{1,1,1}$ has a non-trivial two-cycle. The four-form flux of eleven-dimensional supergravity supported on this cycle adds a vector field to the consistently truncated four-dimensional gauged supergravity. Such extra vector fields are called ${\it Betti}$ vector, and they correspond to {\it baryonic} global symmetries in the dual field theory. Therefore, in addition to the usual mesonic flavor symmetries, baryonic symmetries can mix in the trial R-charge. Furthermore, baryonic flux can be also incorporated in a topological twist, when we consider twisting {\it e.g.} on $\Sigma_g$.
It leads to a puzzle which was addressed in \cite{Azzurli:2017kxo, Hosseini:2019use}:
On the gravity side, there exists an explicit black hole near-horizon solution with baryonic flux \footnote{Black brane solutions with baryonic charges were first studied in \cite{Klebanov:2010tj}.}. However, on the dual field theory side, the contributions of the baryonic charge do not appear in the large-$N$ limit of the free energy \cite{Jafferis:2011zi} and the twisted index \cite{Hosseini:2016tor, Hosseini:2016ume}. 

Another characteristic feature of $M^{1,1,1}$ is that the dual field theory proposal in \cite{Martelli:2008si} is {\it chiral}. 
Based on the development of the localization technique \cite{Pestun:2007rz, Kapustin:2009kz, Jafferis:2010un, Hama:2010av}, many duality checks for ${\cal N}=2$ theories were done by calculating the large-$N$ free energy
\cite{Martelli:2011qj, Cheon:2011vi, Jafferis:2011zi}. Using the matrix model computation \cite{Herzog:2010hf}, field theory calculation showed perfect agreement with the gravity free energy. However, for chiral theories, where the theory is not invariant under conjugation of gauge symmetry representation, it has been known that the matrix model prescription for large-$N$ limit is not applicable since the long-range forces between the eigenvalues do not cancel \cite{Jafferis:2011zi}.
Hence, a new method to calculate the large-$N$ free energy was proposed in \cite{Gulotta:2011si, Gulotta:2011aa}, which uses  counting of gauge invariant operators.  This operator counting method  can be applied to both non-chiral and chiral models. In this paper, we use the operator counting method to study the field theory dual to $M^{1,1,1}$.

The paper is organized as follows. In section \ref{ads/cft}, we review the three-dimensional Chern-Simons-matter theory dual to AdS$_4 \times M^{1,1,1}$ and the operator counting method to calculate the large-$N$ three-sphere free energy. Then, as a warm-up, we employ the volume minimization method and reproduce the field theory result. Section \ref{MV} is devoted to construct the master volume from the toric data of $M^{1,1,1}$. Based on the result of \cite{Gauntlett:2018dpc, Hosseini:2019use}, we calculate the entropy functional from the master volume. We illustrate the procedure using two special cases, instead of studying the most general case. In section \ref{mesonic-sol}, we study the case with the non-trivial mesonic and vanishing baryonic fluxes. We compute the large-$N$ topologically twisted index from the three-sphere free energy. Solving the constraint equations of the master volume, we show that the entropy function matches the twisted index. In section \ref{baryonic-sol}, we first review the black hole solutions charged under the Betti vector field, which corresponds to a baryonic symmetry in field theory. We then calculate the black hole entropy using two different methods, {\it i.e.} using the explicit solution and the variational principles, and show a perfect agreement. We conclude in section \ref{discussion}.
\\

\textbf{Note added:}
While we were finalizing this work, two papers \cite{Hosseini:2019ddy, Gauntlett:2019roi} appeared on the arXiv, with significant overlap with this article. The entropy functional with the mesonic and baryonic magnetic flux, which we study in section \ref{EF-mesonic} and \ref{EF-baryonic},  was discussed in section 5.5 of \cite{Hosseini:2019ddy} and section 4.4 of \cite{Gauntlett:2019roi}, respectively.
\section{AdS/CFT correspondence on AdS$_4 \times M^{1,1,1}$}\label{ads/cft}
In this section, we review the large-$N$ free energy calculation using the operator counting method for the field theory dual of $M^{1,1,1}$ \cite{Gulotta:2011aa}. One of the advantages of the operator counting method is that it also gives  the volume of the five-cycles in the dual seven-dimensional manifold.
This plays an important role in identifying the mixing of baryonic symmetries in the trial R-symmetry.
Then, we turn to the toric geometry for a seven-dimensional Sasaki-Einstein manifold $M^{1,1,1}$.
Using the volume minimization studied in \cite{Martelli:2005tp}, we compute the volume function in terms of Reeb vector. We find the relation between the R-charges of the field theory and the Reeb vector of the gravity theory, and show that the volume computation exactly matches the field theory result off-shell. It implies that F-maximization is equivalent to volume minimization.
\subsection{Operator countings}\label{OPC}
The three-dimensional Chern-Simons-matter theory dual to AdS$_4 \times M^{1,1,1}$ was proposed in \cite{Martelli:2008si}. 
The theory consists of gauge group $U(N)\times U(N)\times U(N)$ and nine bi-fundamental fields $(A_{12,i}, A_{23,i}, A_{31,i})$ with superpotential $W=\epsilon_{ijk}\textrm{Tr}A_{12}^i A_{23}^j A_{31}^k$ and Chern-Simons levels $(2k,-k,-k)$. Here $i=1,2,3.$ As a first check of this duality, they showed that the vacuum moduli space of Chern-Simons theory coincides with a toric CY 4-fold, i.e. a cone over $M^{1,1,1}$ mod ${\mathbb Z}_k$. For comparison of supergravity solutions we will set $k=1$, although field theory side computations can go through for $k\neq 1$ as well. 

\begin{figure}
\centering
\includegraphics[scale=0.4]{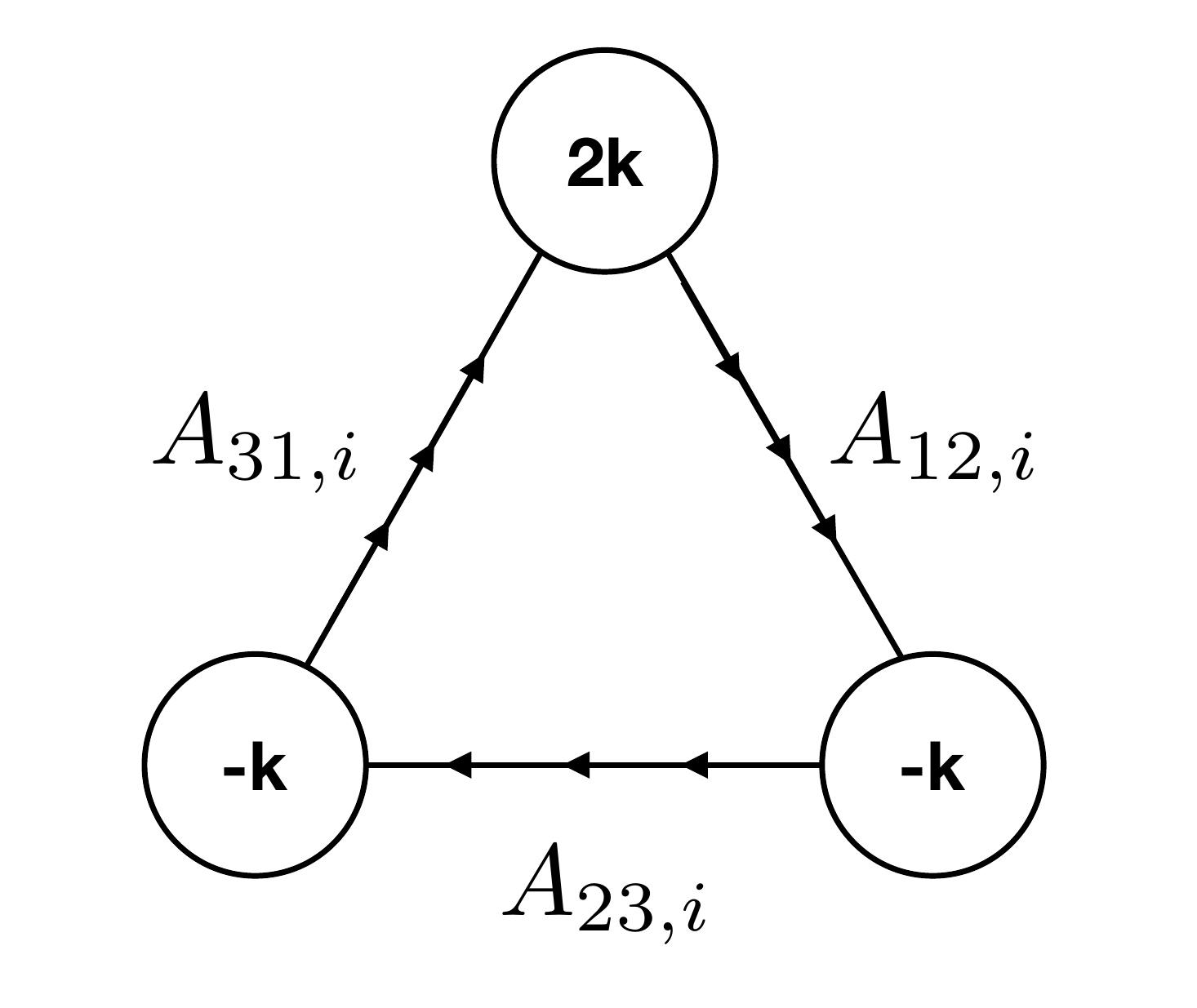}
\caption{The quiver diagram for CS theory dual to AdS$_4 \times M^{1,1,1}$.}
\end{figure}

Let us give a brief summary of the operator counting method developed in \cite{Gulotta:2011si, Gulotta:2011aa}.
The authors counted the number of gauge invariant operators whose R-charges and monopole charges are less than $r, m$, respectively, and devised a new way to obtain the matrix model information such as the eigenvalue density from the number $\psi$  as

\begin{align}
\dfrac{\partial^3 \psi}{\partial r^2 \partial m}\bigg |_{m=r x/\mu}&= \dfrac{r}{\mu} \rho(x),\label{opcrho} \\
\dfrac{\partial^2\psi_{X_{ab}}}{\partial r\partial m}\bigg |_{m=r x/\mu} &= \dfrac{r}{\mu} \rho(x) [y_b (x)- y_a(x)+R(X_{ab})].\label{opccy}
\end{align}
Here $\psi_{X_{a,b}}$ is the number of gauge invariant operators which do not contain bi-fundamental field $X_{a,b}$. 
Now one can calculate the volume of the seven-dimensional internal space and the five-cycles in the dual geometry  as
\begin{align}
\textrm{Vol}(Y_7)&=\frac{\pi^4}{24} \int d \hat{x} \hat{\rho}(\hat{x}). \label{7-vol}\\
\textrm{Vol}(\Sigma_{X_{ab}})&=\frac{\pi^3}{4} \int  d\hat{x} \hat{\rho}(\hat{x})(R[X]+\hat{y_b}(\hat{x})-\hat{y_a}(\hat{x})), \label{cycle-vol}
\end{align}
where
\be
\hat{x} =\frac{x}{\mu},\quad \hat{\rho}(\hat{x})= \frac{\rho{(x)}}{\mu},\quad \hat{y_a}(\hat{x})=y_a(x).
\ee

In this section, we re-visit the operator counting computation done in \cite{Gulotta:2011aa} for the theory dual to $M^{1,1,1}$.
First, let us identify all the $U(1)$ symmetries in the theory.
Having in mind that the isometries of $M^{1,1,1}$ are $SU(3)\times SU(2)\times U(1)$, we have two $U(1)$, namely $U(1)_{1,2}$ symmetries, which are Cartans of $SU(3)$,  $U(1)_3$ symmetry which is a Cartan of SU(2), and also $U(1)$ R-symmetry. Additionally there is $U(1)_\textrm{B}$ which is the baryonic symmetry. Let us summarize the charge assignment as follows.
\begin{center}
\begin{tabular}{c|c c c c c}
      & $U(1)_1$ & $U(1)_2$ & $R_0$  & $U(1)_3$ & $U(1)_B$\\
      \hline
$A_{12,1}$ &     1    &     1    &  2/3 &     1    &     1   \\
$A_{12,2}$ &    -1    &     1    &  2/3 &     1    &     1   \\
$A_{12,3}$ &     0    &    -2    &  2/3 &     1    &     1   \\
\hline
$A_{23,1}$ &     1    &     1    &  2/3 &     0    &    -2   \\
$A_{23,2}$ &    -1    &     1    &  2/3 &     0    &    -2   \\
$A_{23,3}$ &     0    &    -2    &  2/3 &     0    &    -2   \\
\hline
$A_{31,1}$ &     1    &     1    &  2/3 &    -1    &     1   \\
$A_{31,2}$ &    -1    &     1    &  2/3 &    -1    &     1   \\
$A_{31,3}$ &     0    &    -2    &  2/3 &    -1    &     1   \\
\end{tabular}
\end{center}
In general, the trial R-charge of field is a linear combination of all $U(1)$ charges, for example,
\be
R[A_{12,1}]=\dfrac{2}{3}+ \delta_1 +\delta_2+\delta_3+\delta_B.
\ee
However, the free energy functional has many flat directions and is invariant under 
\be
R[X_{a,b}] \rightarrow R[X_{a,b}]+\delta^{(a)}-\delta^{(b)},
\ee
for a bi-fundamental field transforming in the $(\bar{\mathbf{N}},\mathbf{N})$ representation of $U(N)_a \times U(N)_b$ \cite{Jafferis:2011zi}. Using the flat directions, one can set 
\be\label{R-A}
R[A_{12,i}]=R[A_{23,i}]=R[A_{31,i}]=\Delta_i
\ee
 without loss of generality.
Hence, the authors of \cite{Gulotta:2011aa} studied the operator countings with respect to $U(1)_{1,2}$ and R-symmetries. As a consequence of the existence of flat directions, the baryonic symmetry does not contribute to the large-$N$ $S^3$ free energy. It is known that this also happens in the computations of the twisted index \cite{Azzurli:2017kxo}.
Gauge invariant operators can be constructed as 
\be
\begin{array}{lll}
T^{(m)} A_{12}^{2 m k+s}A_{23}^{m k+s}A_{31}^s,&\quad \textrm{for}\,& m>0,\\
T^{(m)} A_{12}^{s}A_{23}^{-m k+s}A_{31}^{-2 m k+s},&\quad  \textrm{for}\,& m<0,
\end{array}
\ee
where $T^{(m)} $ is the diagonal monopole operator, which turns on the same number of units of tr$F_a$ flux for each gauge group. The monopole operator $T^{(1)} $ has a bare R-charge $\Delta_m$ and a gauge charge $(2k,-k,-k)$. By counting the number of gauge invariant operators and evaluating \eqref{opcrho}-\eqref{cycle-vol}, we obtain the volume of the seven-dimensional space and the five-cycles are given as
\begin{align}
&\textrm{Vol}(Y_7)= \dfrac{\pi^4 \left(\Delta_1\, \Delta_2 +\Delta_2\, \Delta_3 +\Delta_3\, \Delta_1 \right)}{216\, k\, \Delta_1^2\, \Delta_2^2\, \Delta_3^2 }, \label{vol_opc}\\
&\textrm{Vol}(\Sigma_{A_{12,i}})=\textrm{Vol}(\Sigma_{A_{31,i}})=
\dfrac{\pi^3 \left(4(\Delta_1\, \Delta_2 +\Delta_2\, \Delta_3 +\Delta_3\, \Delta_1) + 3\,\Delta_1\, \Delta_2\, \Delta_3 \right) }{216\, k\, \Delta_1^2\, \Delta_2^2\, \Delta_3^2 },\\
&\textrm{Vol}(\Sigma_{A_{23,i}})=
\dfrac{\pi^3 \left(2(\Delta_1\, \Delta_2 +\Delta_2\, \Delta_3 +\Delta_3\, \Delta_1) - 3\,\Delta_1\, \Delta_2\, \Delta_3 \right) }{108\, k\, \Delta_1^2\, \Delta_2^2\, \Delta_3^2 }.
\end{align}
Here we set the R-charge of the monopole operator to zero, $\Delta_m=0$, because the monopole operator does not play a role in this paper. The details of computations can be found in \cite{Gulotta:2011aa}.
Imposing the constraint $\Delta_1+\Delta_2+\Delta_3=2$, the volume function \eqref{vol_opc} is extremized at $\Delta_1=\Delta_2=\Delta_3=\frac{2}{3}$ and gives the correct volume of the $M^{1,1,1}/\mathbb{Z}_k$ as
\be\label{volM111}
\textrm{Vol}(M^{1,1,1}/\mathbb{Z}_k) =\dfrac{9 \pi^4}{128k}.
\ee
By using the volume \eqref{vol_opc}, the large-$N$ limit of the free energy becomes
\begin{align}
\label{F-opc}
F&= N^{3/2} \sqrt{\dfrac{2 \pi^6 }{27 \textrm{Vol}(Y_7)}},\nn\\
&= 4 \pi \ \dfrac{\Delta_1 \Delta_2 \Delta_3}{\sqrt{\Delta_1 \Delta_2+\Delta_2 \Delta_3+\Delta_3 \Delta_1}}N^{3/2} k^{1/2}.
\end{align}
To study the cycles of $M^{1,1,1}$, it is more appropriate to work with the following field redefinition
\be
A_{12,i} \equiv u_i v_1, \quad A_{23,i} \equiv u_i, \quad A_{31,i} \equiv u_i v_2.
\ee
Then the volumes of the five-cycles are
\begin{align}
&\textrm{Vol}(\Sigma_{u_{i}})=
\dfrac{\pi^3 \left(2(\Delta_1\, \Delta_2 +\Delta_2\, \Delta_3 +\Delta_3\, \Delta_1) - 3\,\Delta_1\, \Delta_2\, \Delta_3 \right) }{108\, \Delta_1^2\, \Delta_2^2\, \Delta_3^2 },\\
&\textrm{Vol}(\Sigma_{v_{1}})=\textrm{Vol}(\Sigma_{v_{2}})= \dfrac{\pi^3}{24\, \Delta_1\, \Delta_2\, \Delta_3 }.
\end{align}
At the extremized value, the volumes of these cycles become
\be
\textrm{Vol}(\Sigma_{u_{i}}) = \dfrac{3\pi^3}{16}, \quad \textrm{Vol}(\Sigma_{v_{a}}) = \dfrac{9\pi^3}{64}.
\ee
Compared to the R-charge assignment \eqref{R-A}, the R-charges of field $u_i, v_a$ become
\be\label{R-uv}
R[u_i]=\Delta_i, \quad R[v_a]=0.
\ee
\subsubsection{Baryonic symmetry}\label{bs}
Let us recall the well-known relation between the volume of the five-cycles wrapped by an M5-brane and the dimensions of the corresponding baryonic operators \cite{Fabbri:1999hw} as
\footnote{We are using tilde to denote the R-charges mixed with the baryonic symmetry.}

\be\label{R-cyc-vol}
\tilde{R}=\dfrac{\pi}{6} \dfrac{\textrm{Vol}(\Sigma_5)}{\textrm{Vol}(Y)}.
\ee
Then, we can compute the R-charges of the operators $u_i$ and $v_a$ as
\be
\tilde R[u_i]= \dfrac{2}{3}-\dfrac{\Delta_1\, \Delta_2\, \Delta_3}{(\Delta_1\, \Delta_2 +\Delta_2\, \Delta_3 +\Delta_3\, \Delta_1)}, \quad \tilde R[v_a]= \dfrac{3}{2}\dfrac{\,\Delta_1\, \Delta_2\, \Delta_3}{(\Delta_1\, \Delta_2 +\Delta_2\, \Delta_3 +\Delta_3\, \Delta_1)}.
\ee
At extremized values $\Delta_1=\Delta_2=\Delta_3=\tfrac{2}{3}$, the R-charges become
$\tilde R[u_i]=\tfrac{4}{9},\, \tilde R[v_a]=\tfrac{1}{3}$ 
or equivalently $\tilde R[A_{12,i}]=\tfrac{7}{9},\, \tilde R[A_{23,i}]=\tfrac{4}{9},\, \tilde R[A_{31,i}]=\tfrac{7}{9}$. These values 
are consistent with the results of \cite{Hanany:2008fj,Davey:2009qx},
but different from \eqref{R-A}, \eqref{R-uv}.
This discrepancy is due to the baryonic charge.
Natural assignments of $U(1)_3, U(1)_B$ charges for the fields $u_i, v_a$ should be as follows
\begin{center}
\begin{tabular}{c| c c}
         & $U(1)_3$ & $U(1)_B$\\
      \hline
$u_i$    &     0    &     $-2$  \\
$v_1$    &     1    &     3   \\
$v_2$    &     $-1$    &     3  
\end{tabular}
\end{center}
We then have R-charges mixed with $U(1)_3, U(1)_B$ charges as
\be\label{R-FT}
\tilde R[u_i]=\Delta_i-2\delta_B, \quad \tilde R[v_1]=\delta_3+3\delta_B, \quad \tilde R[v_2]=-\delta_3+3\delta_B.
\ee
Then, we can identify the mixing parameters of $U(1)_3, U(1)_B$ as
\be\label{baryon}
\delta_3=0,\quad \delta_B = \dfrac{1}{2}\dfrac{\,\Delta_1\, \Delta_2\, \Delta_3}{(\Delta_1\, \Delta_2 +\Delta_2\, \Delta_3 +\Delta_3\, \Delta_1)}.
\ee
\subsection{Volume minimization}\label{Vmin}
In this section, we turn to the gravity side and calculate the volume of $M^{1,1,1}$
in the context of volume minimization studied by \cite{Martelli:2005tp}\footnote{Similar calculations were done in \cite{Amariti:2011uw}.}. The authors of this reference considered generic toric Sasaki-Einstein manifold $Y$, as a metric cone over $C(Y)$. Once we are given the toric data, one can study a variational problem in the space of all Sasakian metrics, using the familiar Einstein-Hilbert action. This functional is extremized when $Y$ is Einstein and the action gives the volume of the Sasaki-Einstein manifold.

For concreteness, let us now consider the toric data of a seven-dimensional Sasaki-Einstein manifold $M^{1,1,1}/\mathbb{Z}_k$, which can be found {\it e.g.} in \cite{Martelli:2008rt}
\begin{align}\label{toric-M111}
&w_1=(1,0,0,0), \quad w_2=(1,1,0,0), \quad w_3=(1,0,1,0),\nn\\
&w_4=(1,-1,-1,3k), \quad w_5=(1,0,0,2k).
\end{align} 
These vectors constitute the inward-pointing normal vectors of the facets of a convex polyhedral cone $\mathcal C$ in four dimensions.

Then, according to the general results on toric Sasaki-Einstein manifolds, the volume of the internal manifold $Y_7$ and the five-cycles can be calculated from the Euclidean volume of the facet $\textrm{Vol}(\mathcal F_a)$ as follows.
\footnote{We follow the notation of \cite{Jafferis:2011zi}. See section 7.1.}
\begin{align}
\label{vol7}
&\textrm{Vol}(Y_7) =\dfrac{(2\pi)^4}{b_1} \sum_{a=1}^5 \dfrac{1}{|w_a|} \textrm{Vol}(\mathcal F_a), \\
&\textrm{Vol}(\Sigma_{w_a})=6\,(2\pi)^{3}\dfrac{1}{|w_a|} \textrm{Vol}(\mathcal F_a),
\end{align}
where
\be
\dfrac{1}{|w_a|} \textrm{Vol}(\mathcal F_a)= \dfrac{1}{48}
\dfrac{(w_a,w_1,w_2,w_3)^2}{|(b,w_a,w_1,w_2)(b,w_a,w_1,w_3)(b,w_a,w_2,w_3)|}.
\ee
Here $b$ is the Reeb vector.

Now it is straightforward to substitute the toric data \eqref{toric-M111} into \eqref{vol7}, and obtain an expression of the seven-dimensional volume of $M^{1,1,1}/\mathbb{Z}_k$ as a function of the Reeb vector
\begin{align}\label{vol_mini}
\textrm{Vol}(Y_7)
={\scriptstyle
-\frac{6 \pi ^4 k^3 \left(2\, b_4\, k (b_2+b_3-4)-k^2 (b_2-2 b_3+8) (2 b_2-b_3-8)+b_4^2\right)}
{b_4 (3\, b_2\, k+b_4) (3\, b_3\, k+b_4) \left(b_4-k (b_2-2 b_3+8)\right) \left(b_4+k (2\, b_2-b_3-8)\right) \left(b_4+2 k (b_2+b_3-4)\right)}
}
,
\end{align}
where we used $b_1=4$. This expression is minimized at $b_2=b_3=0, b_4=4k$ and minimum gives the correct volume of $M^{1,1,1}/\mathbb{Z}_k$ in \eqref{volM111}.
At the same time we have the volume of five-cycles and the R-charges of the corresponding field theory operators,
\be
\textrm{Vol}(\Sigma_{w_a})=\left(\dfrac{9\pi^3}{64k},\dfrac{3\pi^3}{16k},\dfrac{3\pi^3}{16k},\dfrac{3\pi^3}{16k},\dfrac{9\pi^3}{64k}\right), \quad
\tilde R_a=\left(\dfrac{1}{3}, \dfrac{4}{9}, \dfrac{4}{9}, \dfrac{4}{9},\dfrac{1}{3}\right)
\ee
The baryonic symmetry, which is associated with a relation
\be
\sum B_a w_a =0,
\ee
is given as $(3,-2,-2,-2,3)$.
Compared to the analysis of section \ref{bs}, we conclude that the toric vectors $w_2, w_3, w_4$ correspond to fields $u_i$, and $w_1, w_5$ should be mapped to $v_1, v_2$. In general, R-charges are functions of $b_2, b_3, b_4$. Let us focus on $\tilde R_1 =\tilde R_5$ case, {\it i.e.} $\tilde R_1 -\tilde R_5= \tfrac{1}{2}(4-b_2-b_3-b_4)=0$. This corresponds to the $\delta_3=0$ case in the field theory, which is related to the suppression of the monopole operator. We calculate the R-charges using \eqref{R-cyc-vol} as
\begin{align}\label{Rt-opc}
\tilde R_1&= \tilde R_5
=-\dfrac{\left(4 +2 b_2-b_3 \right) \left(4 - b_2+2b_3 \right)\left(4 - b_2-b_3 \right)}{12 \left(-16+b_2^2 - b_2 b_3 +b_3 ^2 \right)},\nn\\
\tilde R_2&= \dfrac{1}{6} \left(4+2 b_2-b_3 \right)-\dfrac{2}{3} \tilde R_1, \nn\\
\tilde R_3&= \dfrac{1}{6} \left(4- b_2+2 b_3 \right)-\dfrac{2}{3} \tilde R_1, \nn\\
\tilde R_4&= \dfrac{1}{6} \left(4- b_2-b_3 \right)-\dfrac{2}{3} \tilde R_1.
\end{align}
The combinations $\tilde R_{i+1}+\frac{2}{3}\tilde R_1$ are independent of the baryonic symmetry.
We compare \eqref{Rt-opc} and \eqref{R-FT} and identify the R-charges $\tilde R_a$ to those of the field theory operators $\Delta_i $ as $\tilde R_1 =\tilde R_5 \equiv 3\delta_B$ and $\tilde R_{i+1}\equiv \Delta_i-2 \delta_B$. 
This identification leads to
\be\label{R-opc}
\Delta_1= \dfrac{1}{6} \left(4+2b_2-b_3\right), \quad
\Delta_2= \dfrac{1}{6} \left(4-b_2+2b_3\right), \quad
\Delta_3= \dfrac{1}{6} \left(4-b_2-b_3\right),
\ee
or
\be\label{b-opc}
b_2 = 2\left(-2+2\Delta_1+\Delta_2\right), \quad b_3 = 2\left(-2+\Delta_1+2\Delta_2\right) \, . 
\ee

We insert the relation \eqref{b-opc} into the volume function \eqref{vol_mini}. As a result, the volume \eqref{vol_mini} computed from the volume minimization of the toric geometry exactly reduces to the volume \eqref{vol_opc} obtained by operator countings. 

\section{Geometric dual of $\mathcal{I}$-extremization}\label{MV}
In this section, we calculate the master volume from the toric data of $M^{1,1,1}$ \eqref{toric-M111}. The master volume is a generalization of the Sasakian volume studied in the previous section. For a Sasaki-Einstein manifold, the Sasakian volume function can be calculated after we relax the Einstein condition. If we also allow for a general transverse K\"ahler class, the manifold is no longer Sasakian. Then, the volume of this manifold is called the master volume ${\mathcal V}$ \cite{Gauntlett:2018dpc}, because all the physical quantities we need, {\it i.e.} the black hole entropy and R-charges, can be calculated from it.

In terms of toric geometry, the master volume can be calculated from the volume of the polytope ${\mathcal P}$ as \cite{Gauntlett:2018dpc, Hosseini:2019use}
\be
{\mathcal V}=\dfrac{(2\pi)^4}{|\vec{b}|} \textrm {Vol} \left({\mathcal P}(\vec{b} ; \{\lambda_a\}) \right).
\ee
The polytope ${\mathcal P}$ is an intersection of the Reeb hyperplane and the cone $\mathcal C$
\be
{\mathcal P}(\vec{b} ; \{\lambda_a\}) 
\equiv \{ \vec{y} \in H(\vec{b})\,|\,(\vec{y}-\vec{y}_0,\vec{w}_a) \geq \lambda_a, \quad a=1, \cdots,5\},
\ee
where the Reeb hyperplane $H(\vec{b})$ and the origin of the polytope $\vec{y}_0$ are
\be
H(\vec{b}) \equiv \{\vec{y} \in \mathbb{R}^4 \,|\, (\vec{y},\vec{b})=\frac{1}{2}\},
\quad
\vec{y}_0= \left(\dfrac{1}{2b_1},0,0,0\right) \in H.
\ee
Here $\lambda_a$ parametrizes the transverse K\"ahler class.
When we have $\lambda_a=-\dfrac{1}{2b_1}$, the formulae reduce to the Sasakian case.

 The vertex $\vec{y}_I$ of the polytope can be found by solving the equations
\be\label{vertex}
\left(\vec{y}_I-\vec{y}_0,\vec{w}_a \right)=\lambda_a,\quad
\left(\vec{y}_I-\vec{y}_0,\vec{w}_J \right)=\lambda_J,\quad
\left(\vec{y}_I-\vec{y}_0,\vec{w}_K \right)=\lambda_K, \quad
(\vec{y}_I-\vec{y}_0,\vec{b} )=0.
\ee
For a Sasakian polytope, the vertices of the polytope are located at the intersection points of the Reeb hyperplane and the edges of the cone ${\mathcal C}$. The edges are the intersections between three hyperplanes. And each hyperplane is defined to be normal to each toric vector, {\it e.g.} $w_a, w_J, w_K$. Therefore $\vec{y}_I$ is orthogonal to these the toric vectors. When we vary the transverse K\"ahler class so that $\lambda_a \neq -\tfrac{1}{2b_1}$, the vector $\vec{y}_I$ is no longer orthogonal to toric vectors. It implies that the vertices of the Sasakian polytope can be moved by changing the transverse K\"ahler class.
We find the solution to the equations  \eqref{vertex} as
\be
\vec{y}_I=\vec{y}_0 -\dfrac{\lambda_a \left(\vec{b} \times \vec{w}_J \times \vec{w}_K  \right)
                           +\lambda_J \left(\vec{b} \times \vec{w}_K \times \vec{w}_a  \right)
                           +\lambda_K \left(\vec{b} \times \vec{w}_a \times \vec{w}_J  \right)
            }{\left(\vec{w}_J, \vec{w}_K, \vec{w}_a, \vec{b} \right)}.
\ee
We can then calculate the volume of the polytope as 
\begin{align}
\textrm {Vol} \left({\mathcal P}(\vec{b} ; \{\lambda_a\}) \right)
&= \left(\vec{A}, \dfrac{\vec{b}}{|\vec{b}|} \right),\\
&=\left(\dfrac{1}{6} \sum_{a=1}^5 \left(\vec{y}_I-\vec{y}_0 \right) \times \left(\vec{y}_J-\vec{y}_0 \right) \times \left(\vec{y}_K-\vec{y}_0 \right)
, \dfrac{\vec{b}}{|\vec{b}|} \right).
\end{align}
Let us consider a vertex $w_a$ of a given toric diagram of $M^{1,1,1}$. When the number of neighboring vertices of $w_a$ is three, the base of the facet is a triangle in the Reeb hyperplane. 
Since the polytope is three-dimensional, we choose the origin of the polytope $\vec{y}_0$ and compute the volume of the tetrahedron whose vertices are located at $(\vec{y}_I, \vec{y}_J, \vec{y}_K, \vec{y}_0)$. On the other hand, for $w_a$ with four neighbors, the base is quadrilateral. We then break the quadrilateral up into triangles and do a computation.

We obtain the master volume for $M^{1,1,1}$, which are cubic in $\lambda_a$. Due to symmetry reasons it turns out ${\mathcal V}$ depends on only two specific linear combinations of $\lambda_a$. One can easily check ${\mathcal V}$ is invariant under
\be
\lambda_a \rightarrow \lambda_a +\sum_{i=2}^4 l_i \left(b_1 w_a^i-b_i \right) \, , 
\ee
for arbitrary $l_2, l_3, l_4$. Using this, one may choose a particular gauge $\lambda_1 =\lambda_2=\lambda_3=0$.  This is a simple generalization of the $n=3$ case studied in \cite{Hosseini:2019use}. 

Now, let us fibre this seven-dimensional manifold over a two-dimensional Riemann surface $\Sigma_g$. Then, we can study AdS$_2$ $\times$ Y$_9$ solutions of D=11 supergravity theory. This fibration introduces additional parameters $(A, n^i)$.  Here $A$ is a K\"ahler class parameter for the Riemann surface $\Sigma_g$.
The twisting parameter $n^i$ is given by $n^i=\sum_{a=1}^5 w_a^i \mathbf{n}_a$,
where $\mathbf{n}_a$ are the magnetic fluxes, which satisfy the constraint
\be\label{gr-fluxsum}
 \sum_{a=1}^5 \mathbf{n}_a= 2 \left( 1-\mathfrak{g}\right).
\ee
Given the master volume, one has to solve the constraint equation and the flux quantization conditions for $\lambda_a, A$ \cite{Gauntlett:2018dpc}\footnote{We follow the convention of \cite{Hosseini:2019use}. See the equations (4.41)-(4.43) of \cite{Gauntlett:2018dpc} and the equation (3.11), (5.10), (3.14) of \cite{Hosseini:2019use}.} :
\begin{align}\label{eq}
A \sum_{a,b=1}^d \dfrac{\partial^2 {\mathcal V}}{\partial \lambda_a \partial \lambda_b}
&= 2 \pi n^1 \sum_{a=1}^d \dfrac{\partial {\mathcal V}}{\partial \lambda_a}
- 2\pi b_1 \sum_{i=1}^4 n^i \sum_{a=1}^d \dfrac{\partial^2 {\mathcal V}}{\partial \lambda_a \partial b_i},\nn\\
N&= -\sum_{a=1}^d \dfrac{\partial {\mathcal V}}{\partial \lambda_a},\nn\\
\mathbf{n}_a N &= -\dfrac{A}{2\pi }\sum_{b=1}^d \dfrac{\partial^2 {\mathcal V}}{\partial \lambda_a \partial \lambda_b}-b_1 \sum_{i=1}^4 n^i  \dfrac{\partial^2 {\mathcal V}}{\partial \lambda_a \partial b_i}.
\end{align}
Given the solutions, the entropy functional and the R-charges of baryonic operators \cite{Couzens:2018wnk, Hosseini:2019use} are
\begin{align}
\label{ent}
S(b_i, \mathbf{n}_a)&= -8\pi^2 \left (A \sum_{a=1}^d \dfrac{\partial {\mathcal V}}{\partial \lambda_a}
+\left. 2 \pi b_1  \sum_{i=1}^4 n^i \dfrac{\partial {\mathcal V}}{\partial b_i}\right ) \right|_{\lambda_a, A}, \\
\label{Rch}
\tilde{R}_a (b_i,  \mathbf{n}_a) &= -\dfrac{2}{N} \left. \dfrac{\partial {\mathcal V}}{\partial \lambda_a} \right|_{\lambda_a, A}.
\end{align}
The entropy can be obtained by extremizing the functional \eqref{ent} with respect to $b_2, b_3$ and $b_4$ after setting $b_1=1$.

For the $M^{1,1,1}$ case, the expression of the master volume function is quite messy, hence it is not easy to solve \eqref{eq} for $A$ and the two independent $\lambda_a$ directly.  Instead, we will focus on two special cases which allow explicit calculations, in the following.
\section{Case I : black holes with mesonic magnetic flux}\label{mesonic-sol}
In a Sasaki-Einstein manifold, the Reeb vector is Killing and dual to the superconformal $U(1)$ R-symmetry of the dual field theory. Extra isometries of SE manifold are dual to what we call {\it mesonic} flavor symmetries. If there is a non-trivial cycle in SE manifold, the dual field theory has additional symmetry from the reduction of four-form field and we call them {\it baryonic} symmetries.
For the case of $M^{1,1,1}$, its second Betti number $b_2 \left( M^{1,1,1}\right )=1$, {\it i.e.} there is one non-trivial two-cycle. Hence we do have a baryonic global symmetry on field theory side, and its dual gauge field on the gravity side.

First, we calculate the large-$N$ limit of topologically twisted index for $M^{1,1,1}$. Since the chemical potentials and the fluxes associated with the baryonic symmetry do not contribute to the twisted index in the large-$N$ limit \cite{Azzurli:2017kxo}, we first choose to study a simple case, and turn off the baryonic flux. Using the master volume constructed in section \ref{MV}, we calculate the black hole entropy from the toric geometry description and compare with the field theory result. The case where only the baryonic flux is turned on will be studied in the next section.
\subsection{The topologically twisted index}
\label{tti}
In the large-$N$ limit, the topologically twisted index can be expressed in terms of $S^3$ free energy as
\cite{Hosseini:2016tor}
\be
\mathcal{I}\left(\Delta_i, \mathbf{m}_i\right)=\dfrac{1}{2}\sum_{i=1}^3 \mathbf{m}_i \dfrac{\partial F_{S^3} \left(\Delta_i \right)}{\partial \Delta_i},
\ee
where the chemical potential $\Delta_i$ and the fluxes $\mathbf{m}_i$
satisfy the following constraints
\be\label{constraint}
\sum_{i=1}^3 \Delta_i=2, \quad \sum_{i=1}^3  \mathbf{m}_i= 2\left(\mathfrak{g}-1\right)
\footnote{We use $\mathbf{m}_i$ as field theory fluxes and $\mathbf{n}_a$ as supergravity fluxes. The sign convention of sum of fluxes is different to the supergravity case. See \eqref{gr-fluxsum}.}.
\ee
It was shown that such relations hold for non-chiral ${\cal N}=2$ quiver gauge theories \cite{Hosseini:2016tor}. 
    In this section, we assume that this relation can be also extended to {\it chiral} quiver gauge theories, {\it e.g.} gauge theory dual to $M^{1,1,1}$ we are interested in. As we considered in the previous section, the operator counting method provides the $S^3$ free energy \eqref{F-opc}. Using this expression, we can write down the twisted index as
\begin{align}\label{index}
\mathcal{I} \left(\Delta_i,\mathbf{m}_i|
\lambda\right)=& \pi\, N^{3/2} \dfrac{\Delta_1\Delta_2\Delta_3}{\sqrt{\Delta_1\Delta_2+\Delta_2\Delta_3+\Delta_3\Delta_1}}\nn\\
 &\times \Big(\sum_{i=1}^3 \dfrac{\mathbf{m}_i}{\Delta_i}+
\dfrac{\Delta_1\Delta_2\Delta_3}{\Delta_1\Delta_2+\Delta_2\Delta_3+\Delta_3\Delta_1}\sum_{i=1}^3 \dfrac{\mathbf{m}_i}{\Delta_i^2}\Big)\\
&+\pi \lambda (\Delta_1+\Delta_2+\Delta_3-2), \nn
\end{align}
where $\lambda$ is a Lagrange multiplier.
Now we  extremize the index with respect to $\Delta_i$\footnote{We follow here the procedure in \cite{Bobev:2018uxk}.} and solve for $\mathbf{m}_i$.
Inserting these fluxes into the index, it is simply
$\mathcal{I} = - 2 \pi \lambda$.
Plugging the fluxes into the constraint \eqref{constraint}, we determine the Lagrange multiplier in terms of $\Delta_i$.
As the result, the topologically twisted index ${\mathcal I}$ and the magnetic fluxes $\mathbf{m}_i$ can be written in terms of $\Delta_i$. Since the expressions are not particularly illuminating at this moment, we will not write them down here.
Instead, let us consider a simple case where two chemical potentials are set to equal, $\Delta_1=\Delta_3$. Then 
the topologically twisted index reduces to
\begin{align}\label{index-meson}
\mathcal{I}&=\dfrac{8 \pi }{ 3}\left(\mathfrak{g}-1\right)N^{3/2}
\dfrac{\Delta_1^2 \left(\Delta_1^2+6 \Delta_1\Delta_2+3 \Delta_2^2 \right)}
{\sqrt{\Delta_1^2+2 \Delta_1\Delta_2} \left( 4\Delta_1^3+8\Delta_1^2 \Delta_2+4 \Delta_1\Delta_2^2-\Delta_2^3 \right)}.
\end{align}
And the field theory fluxes are written as
\begin{align}\label{flux-meson}
\mathbf{m}_1&=\mathbf{m}_3=\left(\mathfrak{g}-1\right)\dfrac{2 \Delta_1 \left(5 \Delta_1^2+7 \Delta_1\Delta_2+3 \Delta_2^2\right)}{3\left( 4\Delta_1^3+8\Delta_1^2 \Delta_2+4 \Delta_1\Delta_2^2-\Delta_2^3 \right)},\nn\\
\mathbf{m}_2&=\left(\mathfrak{g}-1\right)\dfrac{2\left( 2\Delta_1^3+10\Delta_1^2 \Delta_2+6\Delta_1\Delta_2^2-3\Delta_2^3 \right)}{3\left( 4\Delta_1^3+8\Delta_1^2 \Delta_2+4 \Delta_1\Delta_2^2-\Delta_2^3 \right)}.
\end{align}
Note that, although we here provide expressions for $\mathbf{m}_i$ in terms of $\Delta_i$, conceptually it should be the other way, {\it i.e.} we are after $\Delta_i$ as functions of $\mathbf{m}_i$. Obviously, inverting the above expressions is too cumbersome and we do not attempt to do it.
\subsection{The entropy functional}\label{EF-mesonic}
In this section, we use the master volume $\mathcal V$ computed in section \ref{MV} to show that the topological twisted index in \eqref{index-meson} can be also obtained from toric data.
We restrict ourselves to the simple case considered in section \ref{tti}, and constrain the supergravity fluxes $\mathbf{n}_a$ and the components of the Reeb vector $b_i$ accordingly. We furthermore suppress monopole operators and assume all the fields are uncharged under $U(1)_3$, which implies $ \mathbf{n}_1= \mathbf{n}_5$. 
The symmetry relation $\Delta_1=\Delta_3$ is also translated to $\mathbf{n}_2= \mathbf{n}_4$.
For the Reeb vector components, we assume $b_4=b_1-b_2-b_3$ and $b_2=0$, which are consistent with the expected result $\tilde R_1=\tilde R_5$ and $\tilde R_2=\tilde R_4$.

With these simplifications, we can solve the constraint equation and flux quantization conditions \eqref{eq}. We compute in a particular gauge where $\lambda_1 =\lambda_2 =\lambda_3=0$ and obtain a solution for $(A, \lambda_4, \lambda_5)$. Inserting these solutions into \eqref{ent} and \eqref{Rch}, we obtain the entropy functional and R-charges in terms of $b_i, \mathbf{n}_a$.
The entropy functional is
\be\label{ent-meson-1}
S(b_i, \mathbf{n}_a)=-\dfrac{N^{3/2} \pi \sqrt{-2\left(\sqrt{3} \alpha^{3/2} +9 \alpha \beta-36 \beta^3 \right)}}{ 9b_1 (b_1+2 b_3)^3 \sqrt{b_1^2 \Big((b_1+5b_3)\mathbf{n_3} + 2 \left(b_1+2 b_3 \right)\mathbf{n_4}+ 2 \left(b_1+3 b_3 \right)\mathbf{n_5} \Big)}}
\ee
where
\begin{align}
\alpha &=b_1^2 (b_1+2b_3)^4 \Big\{3\Big((b_1-b_3)\mathbf{n_3}+2(b_1+2b_3)\mathbf{n_4}\Big)^2\nn\\
&\phantom{=}-8(b_1-b_3) \Big((b_1+5b_3)\mathbf{n_3}+2(b_1+2b_3)\mathbf{n_4}\Big)\mathbf{n_5}
-16(b_1-b_3)(b_1+3b_3)\mathbf{n_5}^2\Big\},\nn\\
\beta  &=b_1 (b_1+2 b_3)^2 \Big((b_1-b_3)\mathbf{n_3}+ 2(b_1 +2 b_3)\mathbf{n_4} \Big).
\end{align}
The R-charges are
\begin{align}
\tilde R_1&= \tilde R_5
=\dfrac{-\sqrt{3 \alpha} +3\beta}{6b_1^2 (b_1+2b_3)\Big((b_1+5b_3)\mathbf{n_3} + 2 \left(b_1+2 b_3 \right)\mathbf{n_4}+ 2 \left(b_1+3 b_3 \right)\mathbf{n_5} \Big)},\nn\\
\tilde R_2&= \dfrac{2(b_1-b_3)}{3 b_1}-\dfrac{2}{3} \tilde R_1, \nn\\
\tilde R_3&= \dfrac{2(b_1+2b_3)}{3 b_1}-\dfrac{2}{3} \tilde R_1, \nn\\
\tilde R_4&= \dfrac{2(b_1-b_3)}{3 b_1}-\dfrac{2}{3} \tilde R_1.
\end{align}
First, we identify the chemical potentials in terms of the components of the Reeb vector as
\be\label{map-meson}
\Delta_1= \dfrac{2(b_1-b_3)}{3 b_1}, \quad
\Delta_2= \dfrac{2(b_1+2b_3)}{3 b_1}, \quad
\Delta_3= \dfrac{2(b_1-b_3)}{3 b_1}.
\ee
We also identify $\tilde R_1$ with the baryonic mixing parameter $3\delta_B$, which may include the contribution of the baryonic flux in general. We recall that the mixing parameter $\delta_B$ is not zero even though there is no baryonic flux as it is studied in the section \ref{bs}. Hence, the identification of this baryonic mixing parameter with that obtained in the field theory computation \eqref{baryon} amounts to turning off the baryonic flux.  By using the constraint
$\mathbf{n}_3+2\mathbf{n}_4+2\mathbf{n}_5= 2 \left( 1-\mathfrak{g}\right)$ and plugging the parametrization \eqref{map-meson} into the \eqref{baryon}, this identification leads to
\footnote{It is consistent with the mesonic twist condition $\sum B_a \lambda_a =0$. See the equation (4.17) of \cite{Hosseini:2019ddy}.}
\be
 \mathbf{n}_5=\dfrac{\left(b_1-b_3\right)^2\left( 1-\mathfrak{g}\right)+ 3 b_3 \left(2b_1 +b_3  \right) \mathbf{n}_4 }
 {3b_1^2+2 b_1 b_3+b_3^2}.
\ee
We study the case with the non-zero baryonic flux in the next section where we do not require this identification.  
Second, we identify the supergravity fluxes $\mathbf{n}_1, \mathbf{n}_2, \mathbf{n}_3, \mathbf{n}_4, \mathbf{n}_5$ with the field theory fluxes $\mathbf{m}_1,\mathbf{m}_2,\mathbf{m}_3$ as
\begin{align}
\mathbf{n}_2 +\dfrac{2}{3}\mathbf{n}_5=-\mathbf{m}_1, \quad 
\mathbf{n}_3 +\dfrac{2}{3}\mathbf{n}_5=-\mathbf{m}_2, \quad
\mathbf{n}_4 +\dfrac{2}{3}\mathbf{n}_5=-\mathbf{m}_3.
\end{align}
It leads to the determination $\mathbf{n}_4$ in terms of $b_i$. 
As a result, the supergravity fluxes can be written as
\begin{align}
\mathbf{n}_3&=\left( 1-\mathfrak{g}\right)\dfrac{4\left(b_1 +2b_3 \right) \left( 5 b_1^3-7 b_1^2 b_3 -19 b_1 b_3^2 -6b_3 ^3\right)}{9\left(b_1 +b_3 \right)\left(5 b_1 ^3 - 2b_1^2 b_3 -8 b_1 b_3^2-4 b_3^3 \right)},\nn\\
\mathbf{n}_4&=\left( 1-\mathfrak{g}\right)\dfrac{2\left(b_1 -b_3 \right) \left( 10 b_1^3+ 16 b_1^2 b_3 + 4 b_1 b_3^2 -3b_3 ^3\right)}{9\left(b_1 +b_3 \right)\left(5 b_1 ^3 - 2b_1^2 b_3 -8 b_1 b_3^2-4 b_3^3 \right)},\nn\\
\mathbf{n}_5&=\left( 1-\mathfrak{g}\right)\dfrac{\left(b_1 -b_3 \right) \left( 5 b_1^3+ 8 b_1^2 b_3 + 8 b_1 b_3^2 + 6b_3 ^3\right)}{3\left(b_1 +b_3 \right)\left(5 b_1 ^3 - 2b_1^2 b_3 -8 b_1 b_3^2-4 b_3^3 \right)}.
\end{align}
Inserting the fluxes into \eqref{ent-meson-1} leads to the the final expression of the entropy functional as
\be\label{ent-meson-2}
S(b_i)=\dfrac{8\pi}{9 \sqrt 3} (\mathfrak{g}-1)N^{3/2} \dfrac{\sqrt{b_1}\left(b_1-b_3\right)^2\left(10b_1^2+16b_1 b_3+b_3^2\right)}
{ \sqrt{b_1^2- b_3^2}\left(5b_1^3 -2b_1^2 b_3 -8 b_1 b_3^2 -4 b_3^3 \right)}.
\footnote{Here we assume $\dfrac{-5b_1^2+b_1 b_3+4b_3^2}{-5b_1^3+2b_1^2 b_3+8 b_1 b_3^2+4 b_3^3}<0.$}
\ee
Using the dictionary between the R-charges and components of the Reeb vector \eqref{map-meson}, one can show that the entropy function \eqref{ent-meson-2} exactly match the topologically twisted index \eqref{index-meson} when we set $b_1=1$. 

\section{Case II : black hole with baryonic magnetic fluxes}\label{baryonic-sol}
Now we turn on the baryonic flux only and calculate the black hole entropy. The method developed in \cite{Couzens:2018wnk,Gauntlett:2018dpc} is based on the assumption that there exists a supergravity solution. On the other hand, here our analysis is based on explicit black hole solutions in AdS$_4$.
Let us begin with a consistent truncation of eleven-dimensional supergravity on a seven-dimensional Sasaki-Einstein manifold $M^{1,1,1}$. It is currently not known how to consistently truncate on $M^{1,1,1}$, keeping all the vectors associated with the isometry. At present, the best one can do is to add a {\it Betti} vector associated with the non-trivial two cycle of $M^{1,1,1}$ \cite{Cassani:2012pj}, to the universal SE$_7$ truncation  \cite{Gauntlett:2009zw}.
As a result, the consistently truncated theory is ${\cal{N}}=2$ supergravity with one massless vector multiplet (Betti multiplet), one massive vector multiplet and one hypermultiplet. We note that the massive vector field is associated with a trivial two-cycle. In this theory, AdS$_4$ black hole solutions charged under the Betti vector field was studied numerically in \cite{Halmagyi:2013sla}. Below we review these magnetically charged black hole solutions and calculate the entropy. Then, we use the master volume formula to calculate the entropy from the toric description and show a perfect agreement.

\subsection{The black hole entropy}
In this section, we review the solution studied in \cite{Halmagyi:2013sla} and collect all the information needed to write down the entropy of black holes \footnote{In this section, we explicitly write down the equation number of \cite{Halmagyi:2013sla} for readers' convenience. Here we consider $\kappa=-1$ case only.}.
The near horizon geometry is $AdS_2 \times \Sigma_g$ with radii $R_1$ and $R_2$ respectively for AdS${}_2$ and $S^2$.
The entropy of the black hole is
\be
S_{\textrm{BH}}= \dfrac{\textrm{Area}}{4 G_N^{(4)}}
=\dfrac{128}{9\sqrt 3} \pi N^{3/2} |\mathfrak{g}-1| (R_2)^2,
\ee
where the area of the black hole horizon for $\mathfrak{g} \neq 1$  is
\be
\textrm{Area}=4\pi |\mathfrak{g}-1| (R_2)^2,
\ee
and the four-dimensional Newton's constant is
\be
\dfrac{1}{G_N^{(4)}}=\dfrac{2\sqrt 6 \pi^2 }{9}\dfrac{N^{3/2} }{\sqrt{\textrm{Vol}(Y_7)} R^2} = \dfrac{128}{9\sqrt 3} N^{3/2}.
\ee
Here we used  AdS$_4$ radius as
\be
R=\dfrac{1}{2}\left(\dfrac{e_0}{6}\right)^{3/4},
\ee
with $e_0=6$.
Using the BPS equations, one can write the sphere radius squared $(R_2)^2$ in the language of ${\cal N}=2$ gauged supergravity as \footnote{This quantity as a function of scalar fields is to be extremized. It is called an attractor equation. 
}
\be\label{R2}
(R_2)^2=\mp \dfrac{\text{Re}(e^{-i \psi}\cal{Z})}{{\cal{L}}_i^\Lambda P_\Lambda^3}.
\ee

Let us now explain the quantities in the above expression.
The homogeneous coordinates $X^\Lambda(z^i)$ and a holomorphic prepotential $F(X^\Lambda)$ are needed to describe a special K\"ahler manifold parameterized by the scalars of vector multiplets. The K\"ahler potential is
\be
K= - \textrm{ln}\, i \left(\bar{X}^\Lambda  F_\Lambda -X^\Lambda  \bar{F}_\Lambda\right),
\ee
where $F_\Lambda= \partial_\Lambda F$.
The symplectic section $(L^\Lambda ,M_\Lambda)$ is related to the holomorphic section $(X^\Lambda ,F_\Lambda)$ as
\be
(L^\Lambda ,M_\Lambda)= e^{K/2}(X^\Lambda ,F_\Lambda).
\ee
For the quaternionic manifold parametrized by hypermultiplet scalars, we need the metric and the Killing prepotential.
All the data associated with the model studied in \cite{Halmagyi:2013sla}, {\it i.e.} a prepotential, homogeneous coordinates, the Killing prepotential and the Killing vector are
\begin{align}
F&= 2 \sqrt{X^0 X^1 X^2 X^3}, \quad X^\Lambda=\left(1, z^2 z^3, z^1 z^3, z^1 z^2 \right),\nn \\
P_\Lambda^3&= \sqrt{2} \left(4-\frac{1}{2} e^{2\phi} e_0,-e^{2\phi},-e^{2\phi},-e^{2\phi}\right),
\quad k_\Lambda^a =\sqrt{2} \left(e_0,2,2,2 \right).
\end{align}
Now we can calculate the central charge $\cal{Z}$ and the imaginary part of the symplectic sections ${\cal{L}}_i^\Lambda$ in \eqref{R2} as
\begin{align}
\cal{Z}&= p^\Lambda M_\Lambda-q_\Lambda L^\Lambda, \nn\\
          &=e^{K/2} \Big((p^0 z^1 z^2 z^3 +p^1 z^1+p^2 z^2+p^3 z^3)
          -(q_0+ q_1 z^2 z^3+ q_2 z^3 z^1+ q_3 z^1 z^2\Big),\\
{\cal{L}}_i^\Lambda&= \textrm{Im}\left(e^{-i \psi} L^\Lambda\right), \nn\\
                               &=  \textrm{Im}\Big( -i \dfrac{1}{\sqrt{8 v_1 v_2 v_3}} \left(1, z^2 z^3, z^3 z^1, z^1 z^2\right)\Big).
\end{align}
Here $p^\Lambda, q_\Lambda$ are the magnetic, electric charges, respectively and $z^i= b^i + i v^i$ are the vector multiplet scalars. $\phi$ is one of hypermultiplet scalars.
With the fixed phase $\psi=\pi/2$, the radius squared becomes
\be
(R_2)^2=\mp \dfrac{\text{Re}\Big[ - \dfrac{i}{\sqrt{2}} \Big(p^0 z^1 z^2 z^3 -q_0+
\sum_{a=1}^3 \Big(p^a z^a- q_a \dfrac{z^1 z^2 z^3}{z^a}\Big)\Big)\Big]}
{\text{Im}\Big[-i \Big(4-e^{2\phi}\Big( \dfrac{1}{2}e_0+z^1 z^2+z^2 z^3+z^3 z^1\Big) \Big)\Big]}.
\ee 
We have some constraints on the scalar fields\footnote{See the equations (4.17), (4.18) in \cite{Halmagyi:2013sla}.}
\be
\textrm{Im}(z^1 z^2+z^2 z^3+z^3 z^1)=0 \quad \textrm{and} \quad \textrm{Re}(z^1 z^2+z^2 z^3+z^3 z^1)=-\dfrac{1}{2}e_0.
\ee
With these constraints, the hypermultiplet scalar $e^{2\phi}$ plays a role as a Lagrange multiplier\footnote{This Lagrange multiplier also appears in the study of massive IIA black hole solutions \cite{Benini:2017oxt, Hosseini:2017fjo}.}.
Then the denominator becomes $-4$ and the radius squared becomes
\be
(R_2)^2=\mp \dfrac{\text{Re}\Big[ - \dfrac{i}{\sqrt{2}} \Big(p^0 z^1 z^2 z^3 -q_0+
\sum_{a=1}^3 \Big(p^a z^a- q_a \dfrac{z^1 z^2 z^3}{z^a}\Big)\Big)\Big]}
{-4}
\ee

Now let us focus on $M^{1,1,1}$ case. Following the argument in \cite{Halmagyi:2013sla}, the $M^{1,1,1}$ model can be obtained via equating
\be
b_3= b_1, \quad v_3=v_1, \quad p^3=p^1,\quad q_3=q_1.
\ee
Then the radius squared reduces to
\be
(R_2)^2=\mp \dfrac{\text{Re}\Big[ - \dfrac{i}{\sqrt{2}} \Big(p^0 (z^1)^2 z^2 -q_0+
\Big(2p^1 z^1+p^2 z^2- 2 q_1 z^1 z^2-q_2 (z^1)^2\Big)\Big)\Big]}
{-4}.
\ee
If we insert the solutions $p^0, p^1, p^2, q_1, q_2, b_1, b_2$ (equations (4.49-51), (4.53-54), (4.44-45) in \cite{Halmagyi:2013sla}), we can reproduce equation (4.48) of \cite{Halmagyi:2013sla}
\be
(R_2)^2=\mp\dfrac{v_1^2(2v_1^4+8v_1^3 v_2+(3e_0+8v_1^2)v_2^2)}{16(3 e_0 v_2-4 v_1(v_1+2v_2)^2)}.
\ee
 The radius squared can be written in terms of two real scalar fields $v_1, v_2$. 
\subsubsection{The magnetic black hole entropy}\label{MBH}
 Now let us consider magnetically charged black holes by setting all the electric charge to zero $q_0=q_1=q_2=0$, {\it i.e.} imposing $e_0-2v_1(v_1+2v_2)=0$. Now we set $e_0=6$ and insert $v_2=\dfrac{6-2v_1^2}{4v_1}$ into (4.44)-(4.54) in \cite{Halmagyi:2013sla}. Then we have
\begin{align}\label{m-charge}
& b_1=b_2=0 \quad \textrm{and} \quad 
 p_0=-\dfrac{1}{4\sqrt 2},\, p_1=\dfrac{3+v_1^4}{8\sqrt 2(1+v_1^2)},\, p_2=-\dfrac{v_1^2(-3+v_1^2)}{4\sqrt 2(1+v_1^2)},\\
&(R_1)^2=\dfrac{v_1^2 v_2}{16},\, (R_2)^2= \pm  \dfrac{v_1(9-2 v_1^2+ v_1^4)}{32(1+v_1^2)}\quad \textrm{and} \quad 
e^{2\phi}=\dfrac{8}{9-2v_1^2+v_1^4}.
\end{align}
The black hole entropy becomes
\begin{align}\label{BHE}
S_{\textrm{BH}}&= \dfrac{4\pi}{9\sqrt 3}  \dfrac{v_1(9-2v_1^2+v_1^4)}{1+v_1^2}N^{3/2} |\mathfrak{g}-1|.
\end{align}
Setting $v_1=1$ corresponds to turning off the Betti vector multiplet. It then reduces to the black hole entropy with universal twist
\be
S_{\textrm{BH}}= \dfrac{16\pi}{9\sqrt 3}  N^{3/2} |\mathfrak{g}-1|.
\ee
We will reproduce the black hole entropy \eqref{BHE} using the master volume in the next section.

In \cite{Monten:2016tpu}, the authors write the supergravity vector fields in terms of two massless eigenmodes and a massive one, whose magnetic charges are
\begin{align}\label{BHch}
&P_1= \dfrac{1}{2}(p_0-2p_1-p_2)=-\dfrac{1}{2\sqrt 2},\nn\\
&P_2= \sqrt{\dfrac{2}{3}}(-p_1+p_2)=-\dfrac{\sqrt 3(-1+v_1^2)^2}{8(1+v_1^2)}
,\nn\\
&P_m=\dfrac{1}{2\sqrt 3} (3p_0+2p_1+p_2)=0.
\end{align}
$P_1$ and $P_2$ are the magnetic charges with respect to the graviphoton and the Betti vector field, respectively. $P_m=0$ is consistent with the fact that there is no conserved charge for a massive vector field.
Now we can write down the black hole entropy in terms of the magnetic charges and two real scalars as follows.
\be\label{BHEoff}
S_{\textrm{BH}}= \dfrac{8\pi}{27}  \left(-\sqrt 6 P_1(2v_1+v_2+v_1^2 v_2)+4 P_2(-v_1+v_2) \right) N^{3/2} |\mathfrak{g}-1|.
\ee
This expression will be useful for off-shell matching of the black hole entropy.
\subsection{The entropy functional}\label{EF-baryonic}
Now we study the magnetically charged AdS$_4$ black hole with baryonic flux using the toric geometry description. 
The black hole fluxes correspond to adding baryonic flux to the universal twist.
Since we are interested in the solution which is symmetric under $SU(3) \times SU(2)$, we can set  $\mathbf{n}_1= \mathbf{n}_5$ for SU(2) and $\mathbf{n}_2= \mathbf{n}_3= \mathbf{n}_4$ for SU(3), respectively. For the components of the Reeb vector, we assume $b_4=1-b_2-b_3$ and $b_2=b_3$.

In a gauge where $\lambda_1 =\lambda_2 =\lambda_3=0$, one can easily obtain the solution $(A, \lambda_4, \lambda_5)$. The on-shell value of the master volume is a function of $b_1, b_2, \mathbf{n}_4, \mathbf{n}_5$. We set $b_1=1$ and find that the entropy functional is extremized at $b_2=0$.
 Inserting $b_2=0$, the entropy functional becomes
\be\label{EF}
S=-\dfrac{\sqrt{2} \pi }{9} N^{3/2} \sqrt{\dfrac{9\mathbf{n}_4-\sqrt{81 \mathbf{n}_4^2-72 \mathbf{n}_4\mathbf{n}_5-48\mathbf{n}_5^2}}{3\left(3 \mathbf{n}_4+2 \mathbf{n}_5 \right)}} 
\left( 18\mathbf{n}_4+\sqrt{81 \mathbf{n}_4^2-72 \mathbf{n}_4\mathbf{n}_5-48\mathbf{n}_5^2}\right),
\ee
and the R-charges become
\begin{align}
\tilde{R}_1 &=\tilde{R}_5=3 \times \dfrac{9\mathbf{n}_4-\sqrt{81 \mathbf{n}_4^2-72 \mathbf{n}_4\mathbf{n}_5-48\mathbf{n}_5^2}}{18 (3 \mathbf{n}_4+2 \mathbf{n}_5 )}, \nn\\
\tilde{R}_2 &=\tilde{R}_3=\tilde{R}_4= \dfrac{2}{3}- 2 \times\dfrac{9\mathbf{n}_4-\sqrt{81 \mathbf{n}_4^2-72 \mathbf{n}_4\mathbf{n}_5-48\mathbf{n}_5^2}}{18(3 \mathbf{n}_4+2 \mathbf{n}_5 )}.
\end{align}

When we have $\mathbf{n}_4=\dfrac{4}{9}\left(1-\mathfrak{g}\right), \mathbf{n}_5=\dfrac{1}{3}\left(1-\mathfrak{g}\right)$,  it corresponds to the universal twist without baryonic flux. The R-charges, the on-shell value of the master volume and the entropy functional become
\begin{align}
\tilde{R}_1&=\tilde{R}_5=\dfrac{1}{3}, \quad \tilde{R}_2=\tilde{R}_3=\tilde{R}_4=\dfrac{4}{9},\nn\\
{\mathcal V} &= \dfrac{1}{36\sqrt{3} \pi^2} N^{3/2}, \quad S=\dfrac{16 \pi}{9\sqrt{3}} N^{3/2}\left(\mathfrak{g}-1\right).
\end{align}
It is worth recalling here eq. (5.17) of \cite{Hosseini:2019use}, which relates the $S^3$ partition function and the master volume with universal twist
\be
{\mathcal V}= \dfrac{1}{64 \pi^3} F_{S^3} ,
\ee
One can easily verify it holds indeed here as well. 

Let us now turn to the main task of identifying \eqref{EF} with the black hole entropy \eqref{BHE}.
We need to know how the fluxes $\mathbf{n}_4, \mathbf{n}_5$ and the magnetic charges $p_1, p_2$ are related.
For the universal twist, we already know $3\mathbf{n}_4-4\mathbf{n}_5=0$.
It is then implied that this particular linear combination must be proportional to the baryonic flux, and 
it corresponds to the magnetic charge $P_2 \sim -(p_1-p_2)$ which is coupled to the Betti vector field in \eqref{BHch}.
Additionally, let us consider the constraints  $3\mathbf{n}_4+2\mathbf{n}_5=2\left(1-\mathfrak{g}\right)$ and $2p_1+p_2=\dfrac{3}{4\sqrt{2}}$.
This leads to the identification\footnote{It is more easily seen when we uplift the black hole solution to eleven dimensions.}
\be
\mathbf{n}_4=\dfrac{16\sqrt 2}{9}p_1\left(1-\mathfrak{g}\right), \quad 
\mathbf{n}_5 = \dfrac{4\sqrt 2}{3} p_2\left(1-\mathfrak{g}\right).
\ee
Using the explicit magnetic charge of the black hole \eqref{m-charge}, we obtain
\be
\mathbf{n}_4=\dfrac{4}{9}\left(1-\mathfrak{g}\right)-2B, \quad
\mathbf{n}_5=\dfrac{1}{3}\left(1-\mathfrak{g}\right)+3B , 
\ee
where the baryonic flux $B$ is
\be
B=-\dfrac{\left(1-v_1^2\right)^2}{9\left(1+v_1^2\right)}\left(1-\mathfrak{g}\right).
\ee
Finally we insert these fluxes into the entropy functional \eqref{EF} and obtain exactly\footnote{We consider $\mathfrak{g}>1, v_1 >1$ case.} the black hole entropy \eqref{BHE}.
\section{Discussion}\label{discussion}
In this paper, we have studied the $\mathcal{I}$-extremization and its geometric dual for a seven-dimensional Sasaki-Einstein manifold $M^{1,1,1}$.
The proposed field theory dual to $M^{1,1,1}$ is chiral and it is currently not known how to calculate the large-$N$ limit of the $S^3$ free energy and the twisted index based on the usual matrix model computations. Hence, we have used the operator counting method to compute the $S^3$ free energy and the twisted index at large-$N$. We have checked that this $S^3$ free energy can be obtained from the well-known volume minimization of a seven-dimensional Sasakian manifold. Reproducing this twisted index from the gravity side is one of the main themes of this paper. We have constructed the master volume for $M^{1,1,1}$ by generalizing the Reeb vector and the transverse K\"ahler class. Then, this seven-dimensional space $Y_7$ is no longer Einstein nor Sasakian. We have studied the variational problem defined on the nine-dimensional space which can be obtained by $Y_7$ fibration over a two-dimensional Riemann surface $\Sigma_g$. Turning off the baryonic flux and considering mesonic flux only, we have reproduced the twisted index from the variational principle. Adding a baryonic flux to the universal twist, the variational problem reproduces the entropy of the black hole, which is magnetically charged under the Betti vector field.

At present, our studies have some limitations and we hope to improve the derivation. We note that our computations match {\it on-shell}, {\it i.e.} we have extremized the topologically twisted index and the black hole entropy functional to check agreements. It would be nice to go {\it off-shell} and prove agreements before extremization, as it was possible with ABJM theory \cite{Hosseini:2019use}. In other words, we would like to derive expressions like \eqref{index} and \eqref{BHEoff} from variational principle. The attractor mechanism described in \cite{DallAgata:2010ejj} might give a hint on this.

In this note, we have focused on magnetic index and magnetically charged black hole solutions. On the other hand, dyonic black holes are known to exist \cite{Halmagyi:2013sla}. Generalizations of the twisted index to include electric charges have been studied in \cite{Benini:2016rke, Bobev:2018uxk}. Hence, it is of interest to incorporate also electric charges in the variational problem. A good starting point would be dyonic generalization of the $\mathcal{I}$-extremization and its geometric dual for ABJM theory studied in \cite{Hosseini:2019use}.

On the field theory side, the large-$N$ limit of the $S^3$ free energy and the topologically twisted index do not capture the effect of the baryonic flux \cite{Azzurli:2017kxo}. On the gravity side, the consistently truncated theory including the vector multiplet from the isometry, which corresponds to the mesonic flavor symmetry, is not known. However, the construction studied in this paper includes both mesonic and baryonic fluxes in general.
Due to the complexity of the master volume, we studied here relatively simple cases with either mesonic flux only or baryonic flux only. It would be nice to find more general solutions incorporating both fluxes. 

It would be straightforward to apply the method explained in this paper to other seven-dimensional Sasaki-Einstein manifolds. Another simple example is $Q^{1,1,1}$. Since the second Betti number is $b_2(Q^{1,1,1})=2$, there are two non-trivial two-cycles in the manifold. The near horizon solution of the black hole \cite{Halmagyi:2013sla} and the master volume associated the toric data of $Q^{1,1,1}$ are more complicated than the $M^{1,1,1}$ case.
On the field theory side, the dual Chern-Simons theory is non-chiral, and it is known how to calculate the topologically twisted index \cite{Hosseini:2016ume}.
We can also apply to inhomogeneous Sasaki-Einstein manifolds, for example, $Y^{p,k}\left( \mathbb{CP}^2\right)$ \cite{Martelli:2008rt, Martelli:2008si}. The operator counting method was studied in \cite{Gulotta:2011aa} and successfully reproduced the volume of $Y^{p,k}\left( \mathbb{CP}^2\right)$ \cite{Kim:2012vza} using the AdS/CFT dictionary found in \cite{Kim:2010ja}. We expect that the monopole operator plays an important role in this case. We hope to come back with the answers of these questions in the near future.

\acknowledgments
This work was supported by the National Research Foundation of Korea (NRF) grant NRF-2017R1D1A1B03035515 (HK) and 2018R1D1A1B07045414 (HK, NK).

\appendix
\section{Uplift of $D=4$ solutions with baryonic magnetic charge}
In this appendix, we uplift the near-horizon geometry of the four-dimensional magnetically charged black hole solution \cite{Halmagyi:2013sla}, studied in section \ref{MBH} to eleven dimensions. We show that this uplifted solution is equivalent to the eleven-dimensional solutions \cite{Gauntlett:2019roi, Gauntlett:2006ns, Azzurli:2017kxo} up to overall scaling.

Let us begin with the uplifting formula in \cite{Cassani:2012pj}
\be
ds_{11}^2= e^{2V} \mathcal{K}^{-1}ds_{4}^2+ e^{-V} ds^2(B_6)+e^{2V} (\theta+A^0)^2,
\ee
where $\mathcal{K}$ is related to the K\"ahler potential as $\mathcal{K}=e^{-K}/8$. The left-invariant metric on $M^{1,1,1}$ associated to SU(3) structure is
\be
ds_{7}^2 =\dfrac{1}{8} e^{2U_1} ds^2(\mathbb{CP}^2)+ \dfrac{1}{8}e^{2U_2} ds^2(S^2)+e^{2V}\theta^2.
\ee
Here we set the normalization of $\mathbb{CP}^2$ so that the scalar curvature $R(\mathbb{CP}^2)=4$. The metric can be written in terms of the scalar fields of the vector mutiplets $v_1, v_2$ and the dilaton field $\phi$ with the following field re-definition
\be
v_1 = e^{2U_1+V}, \quad v_2= e^{2U_1+V}, \quad \phi= -2U_1 -U_2.
\ee 
Substituting the near-horizon solutions summarized in the section \ref{MBH} into the uplifting formula, one obtains the eleven-dimensional metric as

\begin{align}\label{uplift}
ds_{11}^2=&\dfrac{1}{4}\left( \dfrac{v_1(3-v_1^2)}{2(9-2v_1^2+v_1^4)}\right)^{2/3}\biggl(ds^2(\textrm{AdS}_2)+16(\theta+A^0)^2 \nn\\
&+(9-2v_1^2+v_1^4)
\left(\dfrac{1}{4v_1^2}ds^2(S^2)+\dfrac{1}{6-2v_1^2}ds^2(\mathbb{CP}^2)+\dfrac{1}{3+2v_1^2-v_1^4} ds^2(\Sigma_g)\right)\biggr).
\end{align}

Now let us consider two eleven-dimensional AdS$_2$ solutions.
In the appendix C of \cite{Gauntlett:2019roi}, the authors discussed AdS$_2 \times Y_9$ solutions found in \cite{Gauntlett:2006ns}. Here $Y_9$ is a U(1) fibration over eight-dimensional K\"ahler space. When the base is the direct product of the K\"ahler-Einstein, the metric, which corresponds to $M^{1,1,1}$, is 
\be\label{GMS}
ds_{11}^2=L^2 e^{-2B/3} \left(ds^2(\textrm{AdS}_2)+(dz+P)^2+e^B
\left(\dfrac{1}{x}ds^2(S^2)+ds^2(\mathbb{CP}^2)+\dfrac{2+x}{1+2x} ds^2(\Sigma_g)\right)\right)
\ee
where 
\be
e^B=\dfrac{3+2x+x^2}{2+x}
\ee
and $x$ is a constant.
If we identify parameters as
\be
v_1\equiv\sqrt{\dfrac{3x}{2+x}},
\ee
the uplifted metric \eqref{uplift} reduces to \eqref{GMS} up to overall factor.

The AdS$_2$ solution with a purely electric baryonic flux studied in section 4.3.1 of \cite{Azzurli:2017kxo} is
\begin{align}\label{ABCMZ}
ds_{11}^2&=L^2 \Bigl( ds^2(\textrm{AdS}_2)+ u\, ds^2(\Sigma_g)+ v\, ds^2(S^2)+ \dfrac{4\,q\,u\,v}{v-u+u\,v}ds^2(\mathbb{CP}^2)\nn\\
&\quad\phantom{L^2 \Big(}+ (d\psi+2qA_B-A+y d\beta)^2\Bigr).
\end{align}
The parameters $u$ and $v$ are not independent and they should satisfy the following constraint (See the equation (4.23) with $w=0$ in \cite{Azzurli:2017kxo}.)
\be\label{constraint-app}
\dfrac{4q^2 \left(-3 \kappa^2 v^2+ 2\kappa u(v-1)v+u^2(v-1)(v+3) \right)}{(\kappa v-uv+u)^4}=0,
\ee
where $\kappa=-1$ for $\Sigma_g$ and $q=R(\mathbb{CP}^2)/8=\frac{1}{2}$. Once we identify
\be
u\equiv\dfrac{3+2x+x^2}{1+2x}, \quad v\equiv\dfrac{3+2x+x^2}{x(2+x)},
\ee
one can easily check that they satisfy the constraint \eqref{constraint-app} and the metric \eqref{ABCMZ} reduces to the metric \eqref{GMS} up to overall factor.

\bibliography{refm111}{}
\end{document}